\documentclass[conference]{IEEEtran}
\makeatletter
\def\ps@headings{%
\def\@oddhead{\mbox{}\scriptsize\rightmark \hfil \thepage}%
\def\@evenhead{\scriptsize\thepage \hfil \leftmark\mbox{}}%
\def\@oddfoot{}%
\def\@evenfoot{}}

\makeatother
\usepackage{graphicx}
\usepackage{times}
\usepackage{graphics}
\usepackage{framed,lipsum}
\usepackage{url}
\usepackage{subfig}
\usepackage{epsfig}
\usepackage{caption}
\usepackage{varwidth}

\usepackage{algorithm}
\usepackage{algorithmicx,algpseudocode}

\newcommand{\CASE}[1]{\STATE \textbf{case} #1\textbf{:} \begin{ALC@g}}
\newcommand{\ENDCASE}{\end{ALC@g}}

\newcommand{\DEFAULT}{\STATE \textbf{default:} \begin{ALC@g}}
\newcommand{\ENDDEFAULT}{\end{ALC@g}}
\newcommand{\DEFAULTLINE}[1]{\STATE \textbf{default:} }
\usepackage{color}
\usepackage{multirow}
\usepackage{multicol}
\usepackage{hhline}
\usepackage{amsthm}
\usepackage{comment}
\usepackage{tabu}

\theoremstyle{plain}

\theoremstyle{definition}

\theoremstyle{remark}

\IEEEoverridecommandlockouts

\begin{document}

\title{Enabling Cooperative IoT Security via Software Defined Networks (SDN)}

\author{
	\centering
\begin{tabular}{c}
		\begin{tabular}{ccc}
			Garegin Grigoryan & Yaoqing Liu & Laurent Njilla \\ grigorg@clarkson.edu & liu@clarkson.edu & laurent.njilla@us.af.mil \\ Clarkson University & Clarkson University & Air Force Research Lab, Rome 
		\end{tabular} \\ \\

			\begin{tabular}{cc}
		Charles Kamhoua & Kevin Kwiat \\ charles.a.kamhoua.civ@mail.mil  & kwiatk@sunyit.edu \\ Army Research Laboratory & SUNY Polytechnic Institute\end{tabular}
\end{tabular}
\vspace{-4mm}
}

\makeatletter
\def\ps@IEEEtitlepagestyle{
	\def\@oddfoot{\mycopyrightnotice}
	\def\@evenfoot{}
}

\def\mycopyrightnotice{
	{\footnotesize
		\begin{minipage}{\textwidth}
			DISTRIBUTION A. Approved for public release: distribution unlimited. Case Number: 88ABW-2017-5015. Dated 16 Oct 2017
		\end{minipage}
	}
}

\maketitle

\pagestyle{empty}

\begin{abstract}

Internet of Things (IoT) is becoming an increasingly attractive target for cybercriminals. We observe that many attacks to IoTs are launched in a collusive way, such as brute-force hacking usernames and passwords, to target at a particular victim. However, most of the time our defending mechanisms to such kind of attacks are carried out individually and independently, which leads to ineffective and weak defense. To this end, we propose to leverage Software Defined Networks (SDN) to enable cooperative security for legacy IP-based IoT devices. SDN decouples control plane and data plane, and can help bridge the knowledge divided between the application and network layers. In this paper, we discuss the IoT security problems and challenges, and present an SDN-based architecture to enable IoT security in a cooperative manner. Furthermore, we implemented a platform that can quickly share the attacking information with peer controllers and block the attacks. We carried out our experiments in both virtual and physical SDN environments with OpenFlow switches. Our evaluation results show that both environments can scale well to handle attacks, but hardware implementation is much more efficient than a virtual one. 


\end{abstract}

\section{Introduction}
\label{sec:intro}

Internet of Things (IoT) is becoming an increasingly attractive target for cybercriminals. We observe that normally IoT systems are static targets and are expected to be attacked due to the fact that many of them have low resource capabilities. They are not equipped with dynamic anti-attack mechanisms to fight against collusive attacks from dedicated attackers~\cite{yu2015handling}. They also work individually and independently to defend from the attacks, which makes themselves to be easily compromised and particularly vulnerable when the same attacking maneuvers can be applied to similar systems since they have the same vulnerabilities. It is important to note, that many illegitimate access attempts are neglected by the current IoT software~\cite{hpstudy}. For instance, we are able to collect a great amount of attacking information from the system logs of an IoT system, when many attackers use a brute-force approach to guess usernames and passwords. Even when these attacks fail, they could be potential threats that may compromise the system sooner or later. However, most of the current practice ignores the failed attack's traces, instead of taking advantage of them. One reason is that such IoT devices have no enough computing capabilities to run custom programs to handle such kind of attacking information. Another more important reason turns out to be the knowledge gap between the application and network layers. The application layer can easily detect application-layer attacks, such as brute-force password guessing, which are transparent to the network layer, because of the original design of TCP/IP protocol stack that naturally separates application-layer and network-layer identifiers. 

To bridge the divide between the application and network layers, Software Defined Networks (SDN) is an emerging and promising technology to make it happen. SDN decouples the control plane from the data plane. An SDN controller can take input from end systems applications and make decisions to the data plane about what traffic can go through. SDN can potentially benefit the security of IoT systems in at least three aspects. First, it can help form a feedback-control loop from end IoT systems to the SDN controller which further controls one or more programmable switches. With the loop, attacks to the victim can be blocked by the intermediate network device, such as an OpenFlow switch. Second, similar attacks to other victims from the same source in the same network can be blocked and thus benefit the whole IoT network in a collaborative way. Moreover, the attacking information can be shared among multiple peering controllers that oversee and control different networks. Other controllers that obtain the attacking information can take the same actions to block the malicious activities before they break into these peering networks. To this end, we designed and built an SDN-based platform that enables cooperative security for IoT systems. We make the following contributions in the paper as follows:

(1) Protocol Design. We define the interfaces, functions and protocols to enable an efficient communication model between victims, OpenFlow switches and the controllers.

(2) Platform Implementation. We implemented the platform using Ryu~\cite{ryu} controllers and OpenFlow switches in GENI~\cite{berman2014geni}. We also tested the platform in a real hardware environment using an OpenFlow switch and a Ryu-based controller.

(3) Performance Evaluation. The platform was evaluated with its effectiveness and efficiency. We measured various times, such as attack detection time, controller response time, controller-to-controller sharing time and flow entry installation time. The evaluation results show that the platform can efficiently share attacking information with trustworthy peers and stop attacks in a timely manner. In addition, hardware-based implementation can handle attacks more efficiently.

The remainder of the paper is structured as follows. Section~\ref{sec:threatmodel} introduces the IoT threat model. Section~\ref{sec:design} describes the overall protocol design and the implementation details. Section~\ref{sec:evaluation} presents the evaluation results. Finally, Section~\ref{sec:conclusion} concludes the paper.
\section{Threat Model}
\label{sec:threatmodel}

A successful cyberattack normally consists of several stages, including reconnaissance, scanning, compromising, escalation. The detection of an attack at its early stages, such as network scanning for discovering open ports, can significantly decrease the chances of the attack escalation. IoT devices are of the particular interest for the attacks that use network scanning, because they have certain ports open all the time (\cite{botnet, IoTissues}). In our threat model, attackers attempt to compromise multiple IoT devices simultaneously in different networks that are connected by a few programmable switches, such as OpenFlow switches. The switches are further controlled by many logically centralized controllers. We also assume that the IoT devices that are under attacks have not been compromised yet. Namely,  we primarily focus on stopping attacks when they are still in scanning and compromising stages. We also posit that some of the IoT devices in the networks are able to detect attacks happening through the information collected by their applications, e.g., system logs. They can extract attackers' identity information, e.g., source IP addresses or port numbers. Also these devices can run small custom programs, such as small SDN agent programs to communicate with the SDN controller. Normally such devices should be powerful enough to run our SDN agent, which requires relatively small CPU resources. Note that our design \textit{does not require} every IoT device to run a user program to identify attacks due to the low resource capabilities of the small devices. However, in our proposed topology those devices will be protected as well thanks to more powerful devices in the network.

\section{Design Overview}
\label{sec:design}

\subsection{Overall Design}
\begin{figure}
	\centering
	\includegraphics[width=3in]{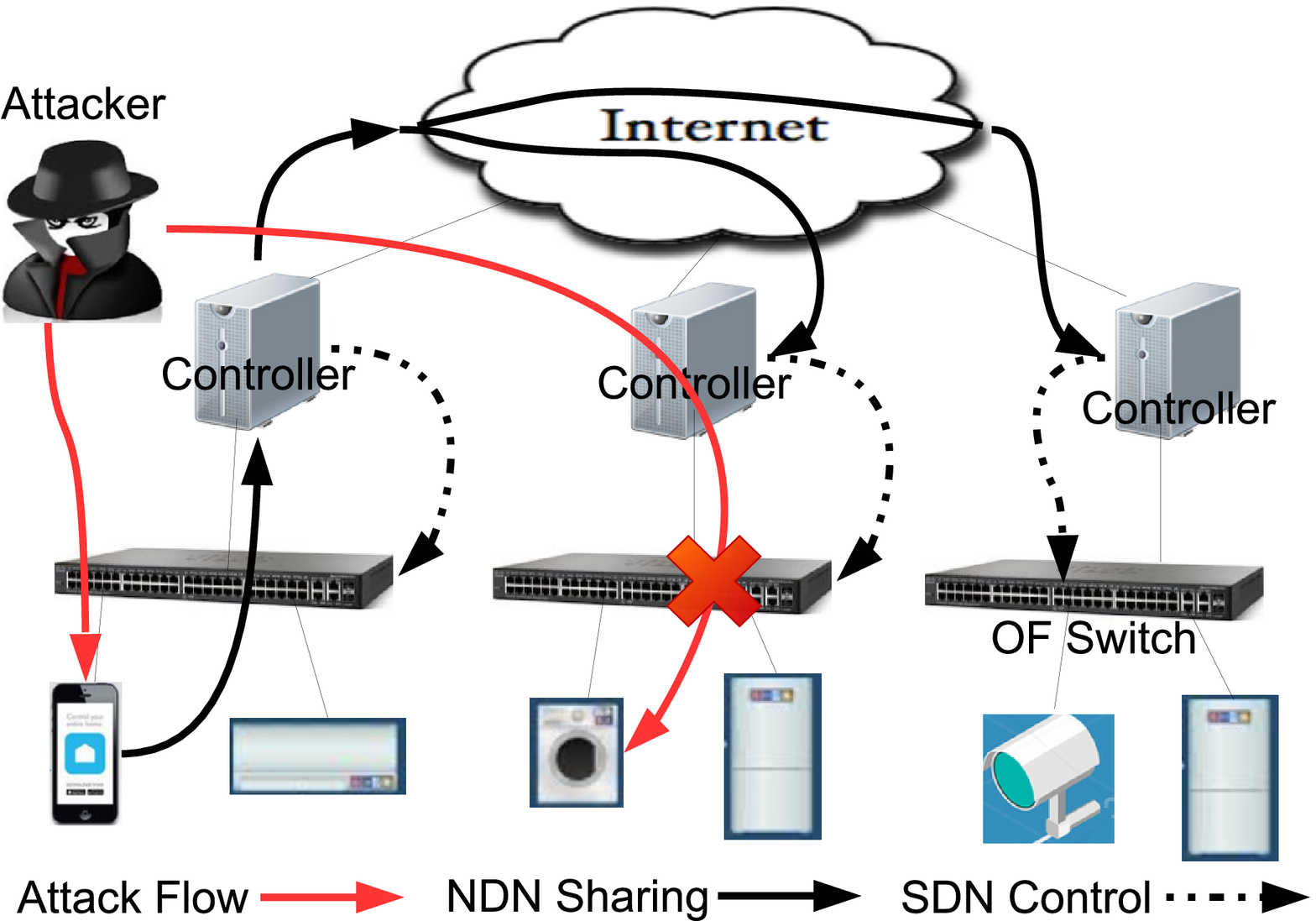}

	\caption{Architecture of Cooperative IoT Security}
	\label{fig:framework}
	\vspace{-8mm}
\end{figure}

In the design, we leverage SDN to build a scalable framework that enables trustworthy and cooperative IoT security. Figure~\ref{fig:framework} illustrates the high-level architecture. When an attacker attempts to attack multiple IoT devices located in different networks, assume one of them can detect such attacks, e.g., a mobile phone. The phone runs a user program to compose a message that contains the attacker's identity, e.g., IP address. The message then is shared with the SDN controller over UDP or TCP. After confirming the victim is a registered IoT system, an SDN application running on the controller is used to compose an OpenFlow switch message. The message is then issued to the OpenFlow programmable switch. Once the command takes effect, the attack will be stopped and other IoT systems in the network will not receive similar malicious attempts any longer. Meanwhile, the attacking information from the victim is shared with other controllers that cover different networks. The other controllers perform the same tasks as the first controller did when they receive the information: issuing controlling commands to their programmable switches to filter out malicious traffic. As we can see from Figure~\ref{fig:framework}, an attack to the washing machine in the second network is blocked because of the cooperation. The following sections present the detailed designs for corresponding interfaces, functions and protocols.

\subsection{Protocol Design}

A controller needs to handle many events, therefore, when we design the protocol between controllers and victims, a few modules are considered in the controller as follows.

(1) Initialization. During boot-up stage, the controller needs to initialize itself and the OpenFlow switch, where a default table-miss flow entry needs to be installed. Because initially the flow table is empty and no data packets can be forwarded between end hosts, the default flow entry can direct all received packets to the controller, which subsequently processes these packets based on their types and content. 

(2) Registration. A controller needs to ensure the information it receives is trustworthy, thus whoever sharing attacking information with it should register themselves at the controller. The IoT systems that have good resource capabilities and can detect attacks need to be registered. Otherwise, a malicious user or system can fool the controller through sharing arbitrary false attacking information. Other controllers that want to be shared with and are willing to share attacking information need to be registered as well to benefit their networks. The controller uses a specific port to receive messages from them and maintains a list of registered IoT systems and controllers that can share information with each other. In addition, as an acknowledgment message, the controller sends a list of required parameters, such as passcode, timestamp, source IP address and port number to the registrars. When an attack occurs, the registered victim needs to fill out the required parameters in a data packet and share them with the controller.

(3) Normal Packet Handling. Two types of normal packets are handled by the controller: Address Resolution Protocol (ARP) packets and IP packets. Once receiving an ARP request message, the controller replies with the MAC address of the unknown IP address to the requester. For an IP packet, which contains source and destination IP addresses, the controller finds the corresponding output port and forwards the packet to that port. This process is not done by the controller for every packet. Instead, the controller installs a flow entry in the flow table of the OpenFlow switch. Subsequently, an IP packet that has the same destination IP address can be routed by the flow rule in the flow table to its next hop. 

(4) Attacking Information Message Handling. One of the most important job for the controller is to handle messages from the registered IoT systems, which as victims under attacks may send the attacking information to the controller for help. Upon receiving such message, the controller needs to confirm that the sender is a registered and trustworthy host. It is easy to confirm a registered host since the host sends to the controller the passcode assigned during the registration. In addition, the host's IP address must be found in the list of IP addresses of all registered hosts that the controller maintains. To confirm a trustworthy host, it is relatively tricky, since the challenge is that how the controller can tell that the host does not fake a source IP address to ask the controller to block it. In other words, the controller needs to see certain evidence to justify that the reported suspicious traffic exists. Recall that any IP packet with previously unknown source or destination IP address needs to be forwarded to the controller by the default table-miss rule installed in the OpenFlow switch, thus the controller keeps a record about the prior communication of the reported attack if it exists. If a victim reports an attack, the controller searches a relevant record. If such record exists, the controller trusts that the attack is ongoing and takes actions to block the attack. This approach may not be able to fully stop a host that has been compromised to report a false alarm, but it is a good level of safeguard to ensure that the host does not misuse its privileges. On the other hand, if the message is from a neighbor controller, we will conduct similar checks. It is possible that the neighbor network has not yet been under attacks by the same source, so the neighbor controller will not install the flow entry immediately to the flow table in the OpenFlow switch. However, it will keep the attacker's information in its repository. When a new data packet, originated from the same source, appears at the controller, it will install the flow rules to the programmable switch to stop the attack. As a result, both the entire victim and neighbor networks can benefit from such information sharing and collaboration. 

\begin{figure}
	\includegraphics[width=3.6in]{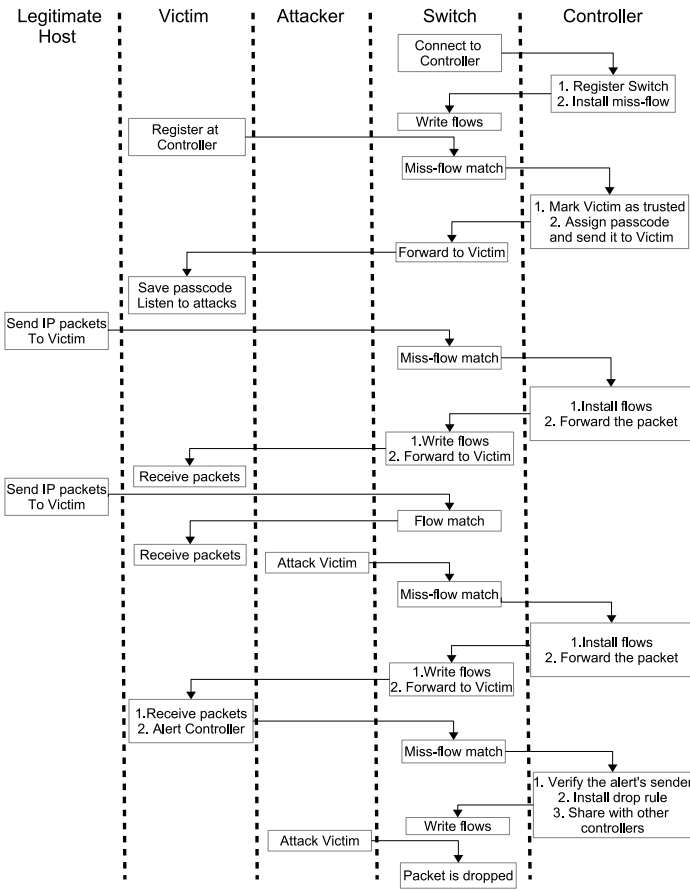}
	
	\caption{SDN Workflow}
	\label{fig:sdnworkflow}
	\vspace{-8mm}
\end{figure}

Figure~\ref{fig:sdnworkflow} shows the workflow between different components. We skip the second controller on the figure due to the limited space. Algorithms~\ref{alg:controller} and~\ref{alg:patriciabuild} show the pseudo code of both controller and victim. 




  \algnewcommand\algorithmicforeach{\textbf{for each}}
 \algdef{S}[FOR]{ForEach}[1]{\algorithmicforeach\ #1\ \algorithmicdo}
\algnewcommand\algorithmicswitch{\textbf{switch}}
\algnewcommand\algorithmiccase{\textbf{case}}
\algnewcommand\algorithmicassert{\texttt{assert}}
\algnewcommand\Assert[1]{\State \algorithmicassert(#1)}%
\algdef{SE}[SWITCH]{Switch}{EndSwitch}[1]{\algorithmicswitch\ #1\ \algorithmicdo}{\algorithmicend\ \algorithmicswitch}%
\algdef{SE}[CASE]{Case}{EndCase}[1]{\algorithmiccase\ #1}{\algorithmicend\ \algorithmiccase}%
\algtext*{EndSwitch}%
\algtext*{EndCase}%
\algdef{SE}[DOWHILE]{Do}{doWhile}{\algorithmicdo}[1]{\algorithmicwhile\ #1}%
\begin{algorithm}
	\caption{Controller's program pseudocode}\label{alg:controller}
	\begin{algorithmic}[1]
		\Procedure{$event\_on\_switch\_connect$}{$s$} \Comment{Triggered when the switch $s$ connects to the controller}
		\State Initialize a hash table $T_s$ that stores the miss-flow matched packet's source IP addresses
		\State Install the miss-flow into $s$
		\State Add $s$ to the list $S$ of connected switches
		\EndProcedure
		\Procedure{$event\_on\_peer\_message$}{$ip$}\Comment{Triggered by an attack alert message from a peer controller}
		\If {Alert's sender is verified}
		\State Add $ip$ to the block list $L$
		\ForEach {$s$ in $S$}
		\If {$ip$ exists in $T_s$}
		\State Add the drop rule for $ip$ on $s$
		\EndIf
		\EndFor
		\EndIf
		\EndProcedure
		\Procedure{$event\_on\_ip\_packet$}{$p$, $s$}\Comment{Triggered by a miss-flow matched packet $p$ on the switch $s$}
		\State Update $T_s$ with $p.ip\_src$
		\Switch{packet $p$}
		\Case {$p$ is a registration packet:}
		\State Register the host with $p.ip\_src$ as a ``trusted" 
		\State Generate and send a passcode  to $p.ip\_src$
		\EndCase
		\Case {$p$ is an attack alert and $p.ip\_src$ is verified as ``trusted":}
		\State Extract attacker's IP address $ip$ from $p$
		\If {$ip$ exists in $T_s$}
		\State Install the drop rule for $ip$ on $s$
		\State Share $ip$ with the peer controllers
		\EndIf
		\EndCase
		\Case {$p$ is an ARP request for $ip$}
		\State Send to $p.ip\_src$ the next hop MAC for $ip$
		\EndCase
		\Case {default:}
		\If {$p.ip\_src$ is in a block list $L$}
		\State Install the drop rule for $p.ip\_src$ on $s$
		\Else 
		\State Install the flow rules for $p.ip\_src$
		\State Forward $p$ to the output port
		\EndIf
		\EndCase
		\EndSwitch
		\EndProcedure
	\end{algorithmic}
\end{algorithm}

\begin{algorithm}
	\caption{Victim's agent pseudocode}\label{alg:patriciabuild}
	\begin{algorithmic}[1]
		\Procedure{$victim\_defense$}{}
		\State Send the registration packet to the controller
		\State Store the passcode assigned by the controller
		\State Listen to attacks
		\If {an attack from $ip$ is detected}
		\State Send the alert message to the controller containing $ip$ and the passcode
		\EndIf
		\EndProcedure
	\end{algorithmic}
\end{algorithm}

\section{Evaluation}
\label{sec:evaluation}

\subsection{Methodology}

\begin{figure}
	\centering
	\includegraphics[width=3in]{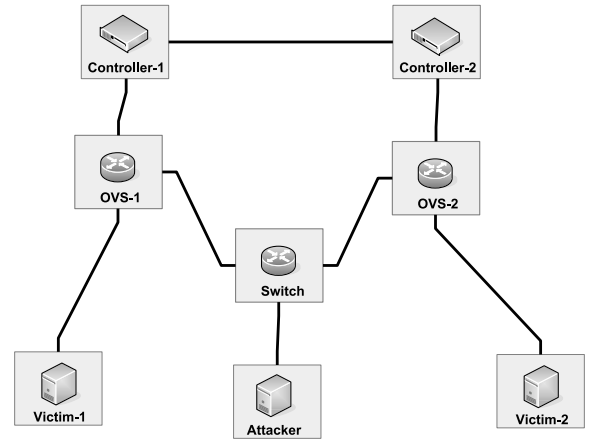}
	\caption{GENI Experimental Topology}
	\label{fig:topology}
	\vspace{-8mm}
\end{figure}

We built up an experimental environment in GENI testbed~\cite{whatisgeni}. GENI (Global Environment for Network Innovations) is an open infrastructure for at-scale networking and distributed systems research and education that spans the US. In the testbed, two controllers, two OpenFlow switches, two victims and one attacker are emulated and located on different Virtual Machines (VMs), as shown in Figure~\ref{fig:topology}. The attacker is connected with two networks which are controlled by the two controllers. Each of the controllers operates one OpenFlow switch on which Open Virtual Switch (OVS) software is running. 
We set up the experiments as follows. Initially, both controllers are initialized and install miss-flow entries in their OpenFlow switches. After that, potential victims communicate with their controllers and the two controllers talk to each other for registration purpose. Eventually, the victims and the peer controllers are added to the list of the trusted hosts. Subsequently, the attacker launches the attack to the first victim and the first packet is re-directed to the first controller. The controller installs corresponding flow entries to the first OVS and lets the following packets from the same source go through. When the attacker's packets arrive at the first victim, the victim immediately discovers that it is an attack. Instantly, the victim extracts the attacker's identity information and sends it to the first controller. After verifying the victim as ``trusted" and that the attacker's data trace was recorded previously, the controller issues OpenFlow commands to stop the attack and meanwhile shares the attacking information with the peer controller, which takes similar actions to block the attack. 
Based on the scenario above, we measured: (1) the time when the attacker's packet reached the controller; (2) the time when the victim detected the attack; (3) the time when the first controller obtained the attacking information; (4) the time when the flow entries were installed in OVS-1; (5) the time when the second controller received the attacking information and (6) the time when the flow entries were installed in OVS-2. We used iperf client tool~\cite{iperf} to simulate the attack.

In addition to the GENI experiment, we conducted similar experiments on a real hardware environment, including a Pica8 OpenFlow switch~\cite{Pica8}, a controller with Intel(R) Core(TM)2 Duo CPU E8400 @ 3.00Ghz machine, an attacker and a victim on two Raspberry Pis~\cite{RaspberryPi} with ARMv7 Processor (rev 4 (v71)). We performed 10 rounds of experiments for each setting.


%


\subsection{Results in GENI environment}


\subsubsection{Attack alert time} 
\begin{figure*}[t]
		\centering
		\captionsetup{justification=centering, width=0.32\linewidth}
		\subfloat[Attack alert time]
		\centering
		\includegraphics[width=0.32\linewidth]{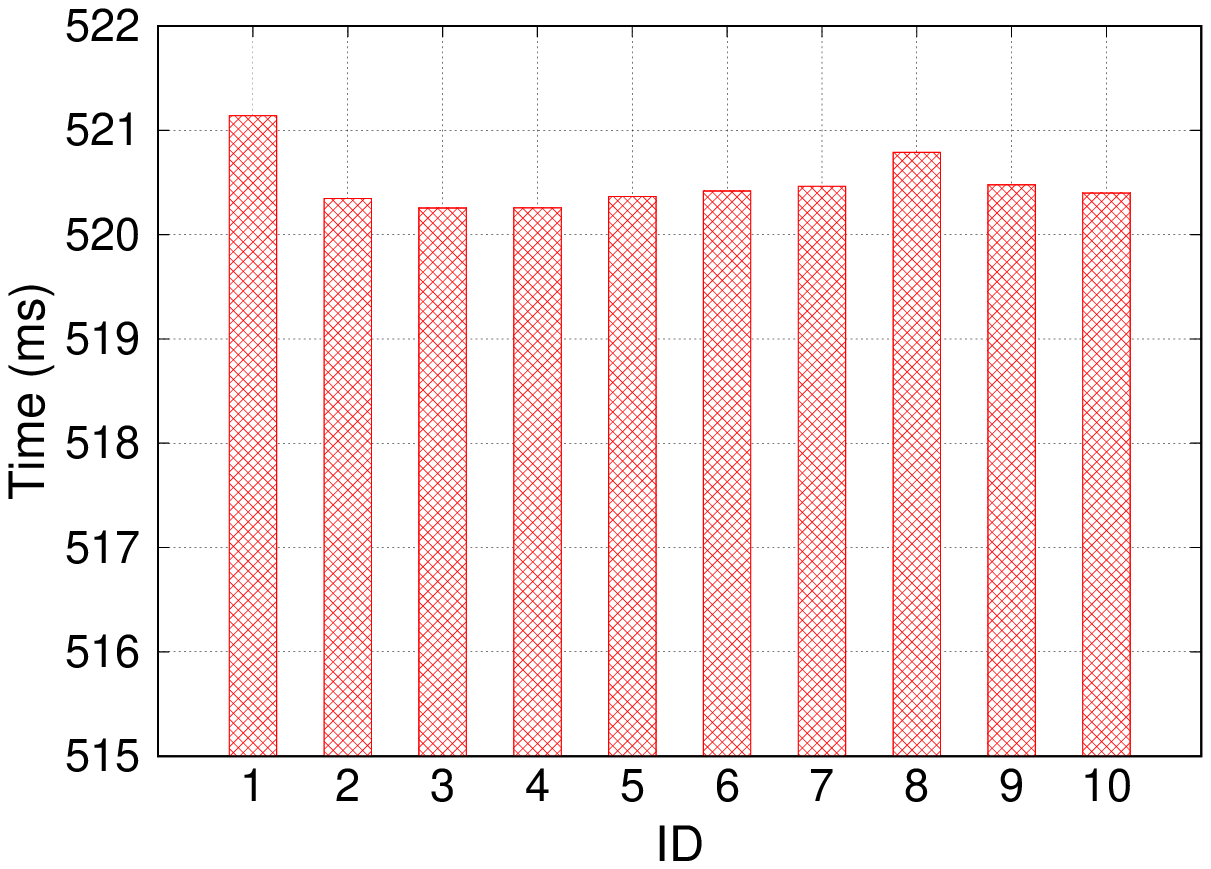}
		\label{fig:VictimAlertTime}
		\subfloat[Flow installation time on OVS-1]
		\centering
		\includegraphics[width=0.32\linewidth]{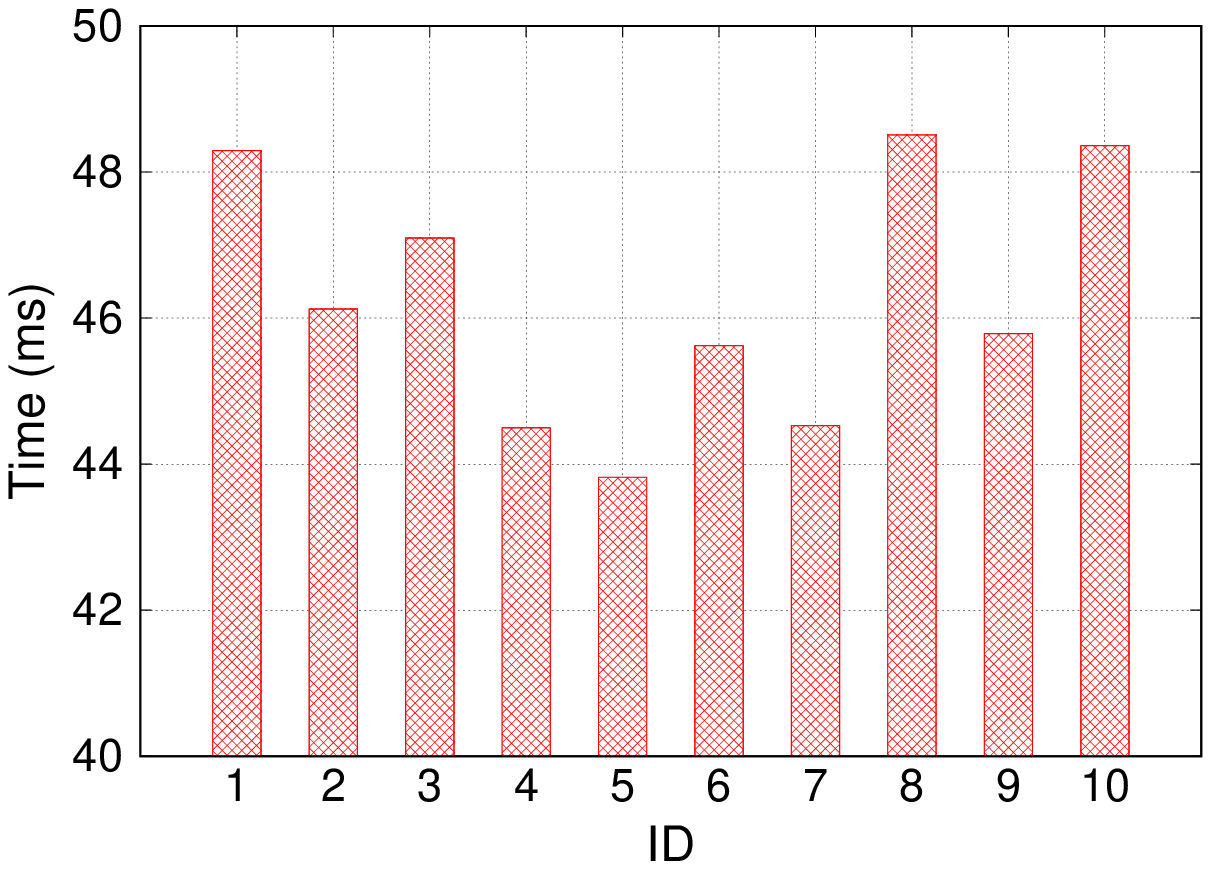}
		\label{fig:flowinstallation1}	
		\subfloat[Total time to defend the network]
		\centering
		\includegraphics[width=0.32\linewidth]{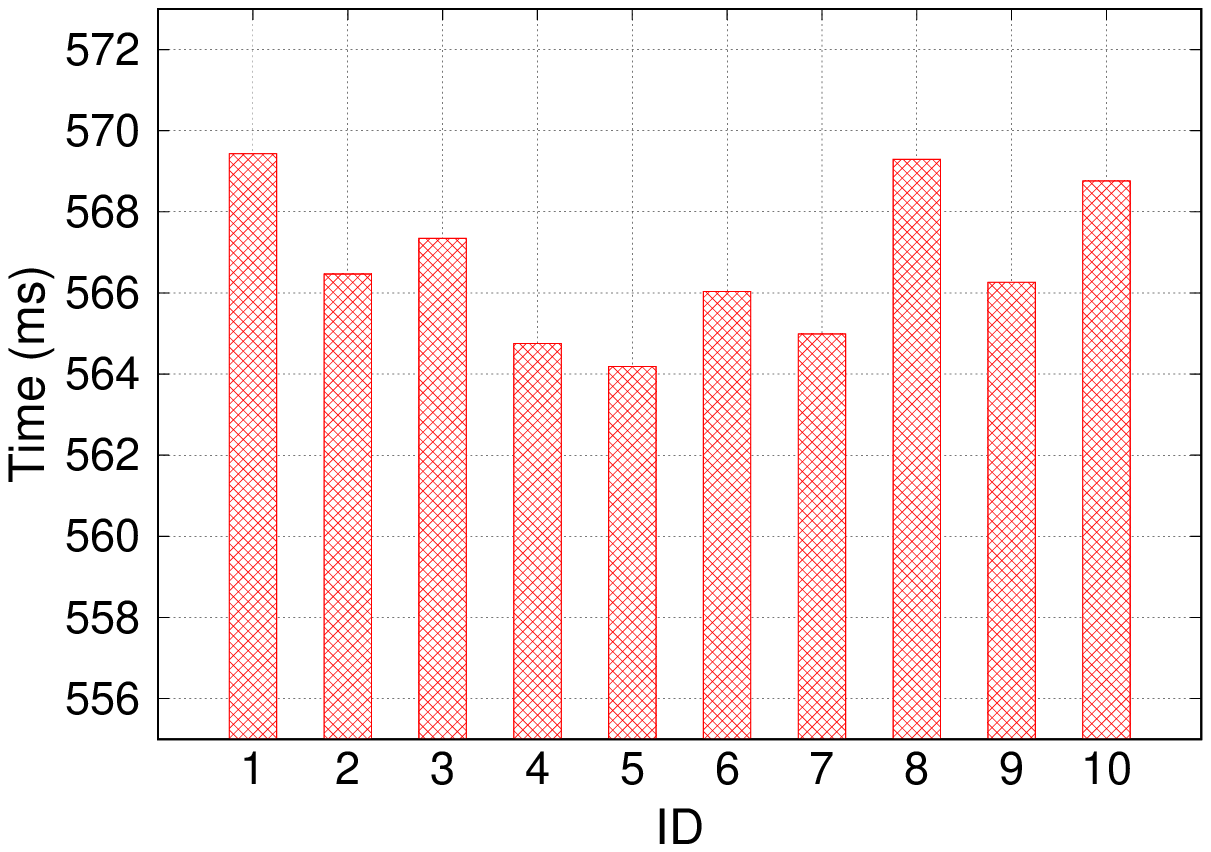}
		\label{fig:halfround}
	\vspace{-2mm}	
		\captionsetup{justification=centering, width=1\linewidth}
	\caption{GENI experiments (Controller-1's network)}
	\label{fig:virtexperiment1}
	\vspace{-4mm}
\end{figure*}

Figure~\ref{fig:virtexperiment1}(a) shows the attack alert time by the victim to the controller. It is the duration from when the attack was discovered by the victim to when the controller received the alert message. On average, it takes around 520 $ms$ to alert a new attack for the ten experiments.


\begin{figure*}[t]

		\centering
		\captionsetup{justification=centering, width=0.32\linewidth}
		\subfloat[Information sharing time between two controllers]
		\centering
		\includegraphics[width=0.32\linewidth]{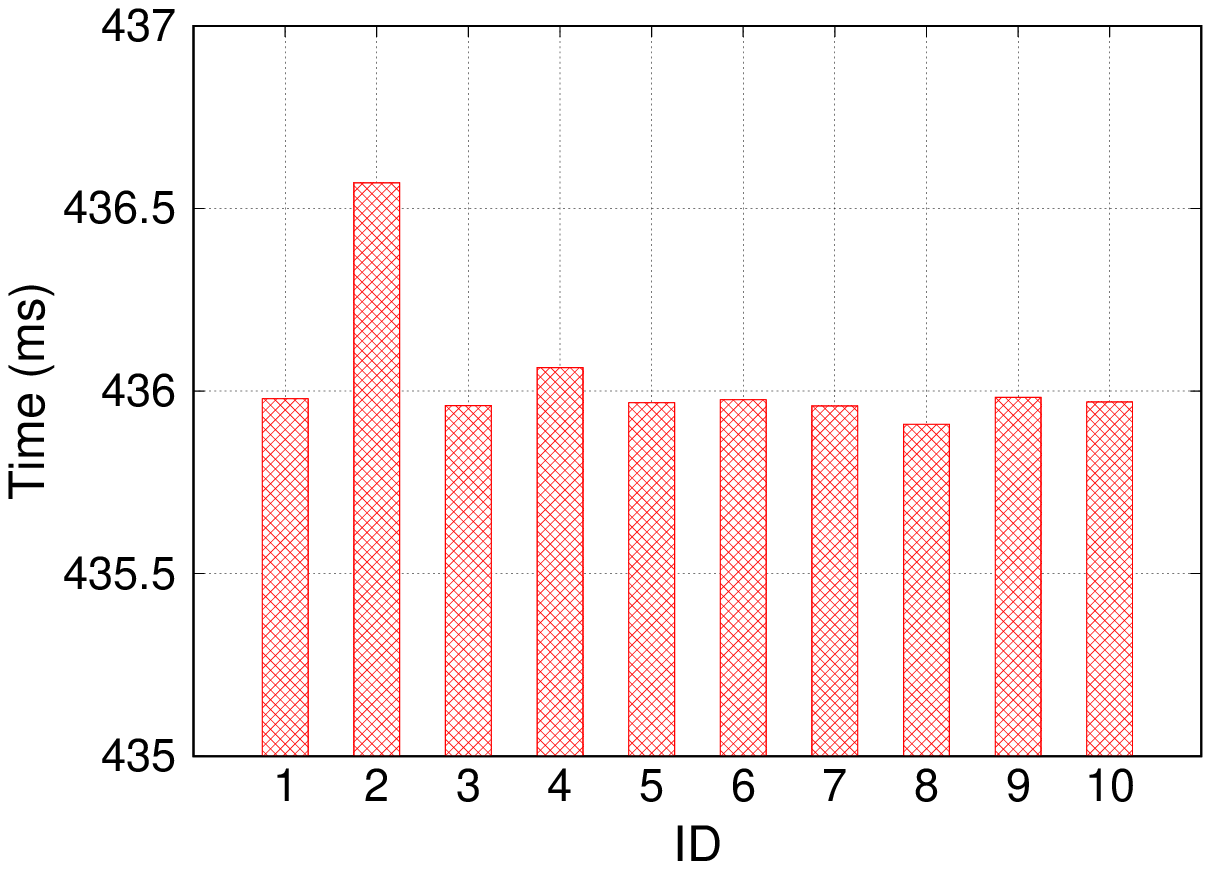}
		\label{fig:sharingtime}
		\subfloat[Flow installation time on OVS-2]
		\centering
		\includegraphics[width=0.32\linewidth]{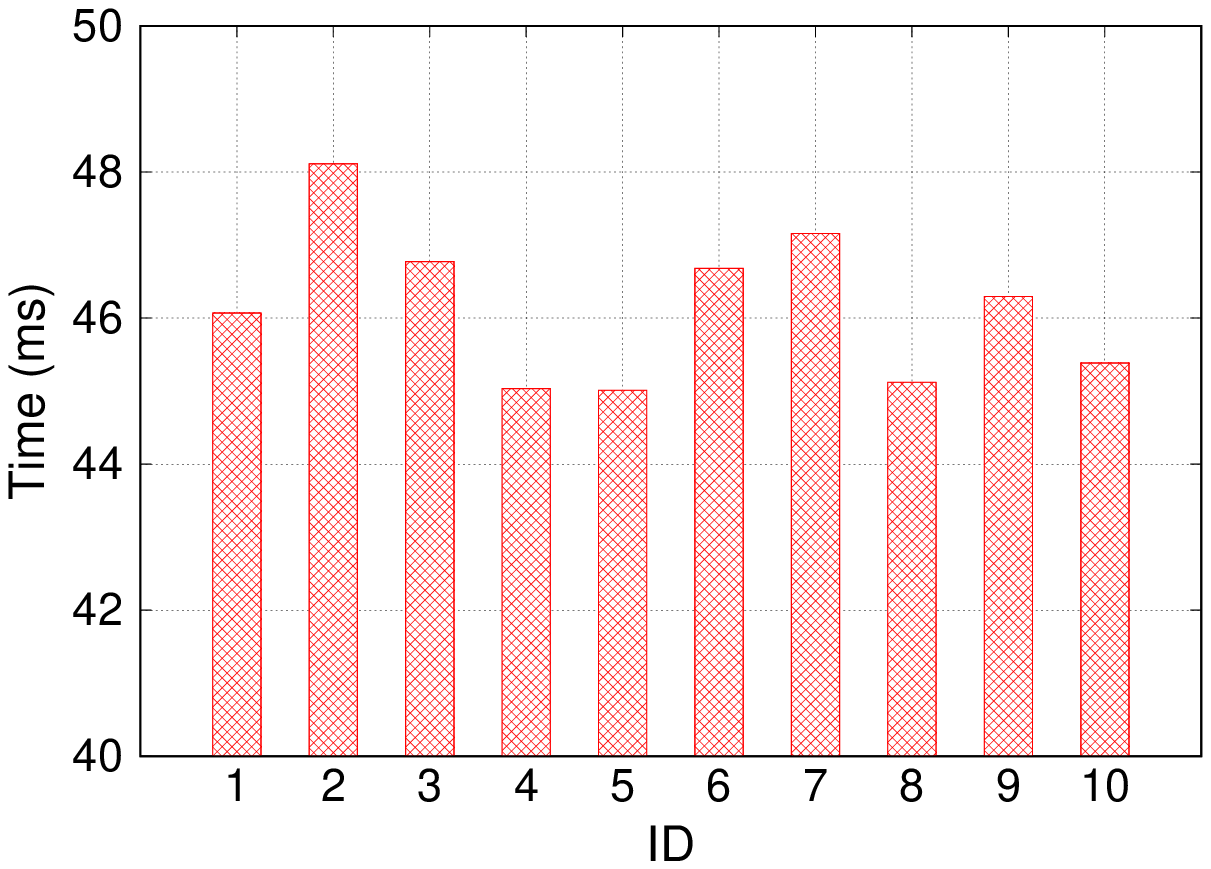}
		\label{fig:flowinstallation2}	
		\subfloat[Total time to defend the second network]
		\centering
		\includegraphics[width=0.32\linewidth]{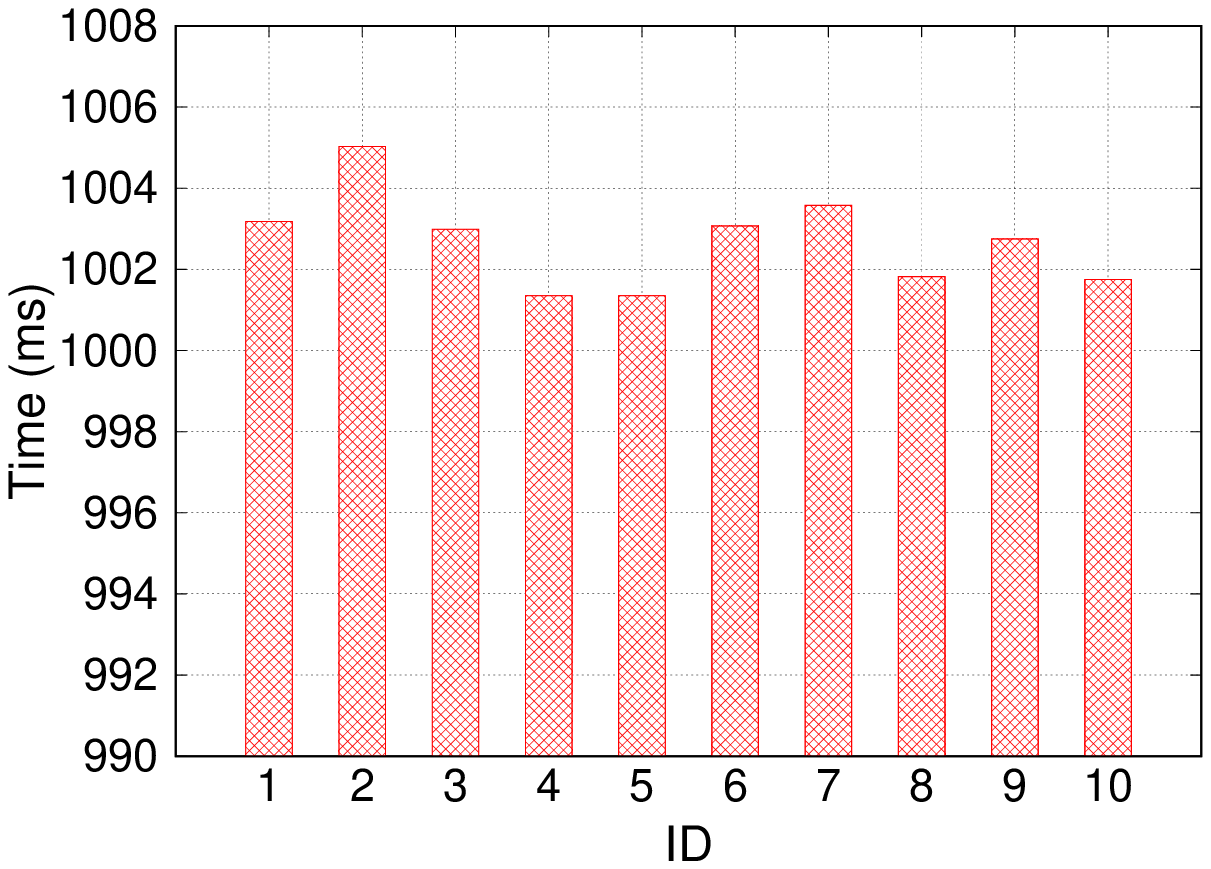}
		\label{fig:totalround}
		
	\vspace{-2mm}	
\captionsetup{justification=centering, width=1\linewidth}
	\caption{GENI experiments (Controller-2's network)}
	\label{fig:virtexperiment2}
	\vspace{-2mm}
\end{figure*}

\begin{figure*}[t]
		\centering
		\captionsetup{justification=centering, width=0.32\linewidth}
		\subfloat[Attack alert time]
		\centering
		\includegraphics[width=0.32\linewidth]{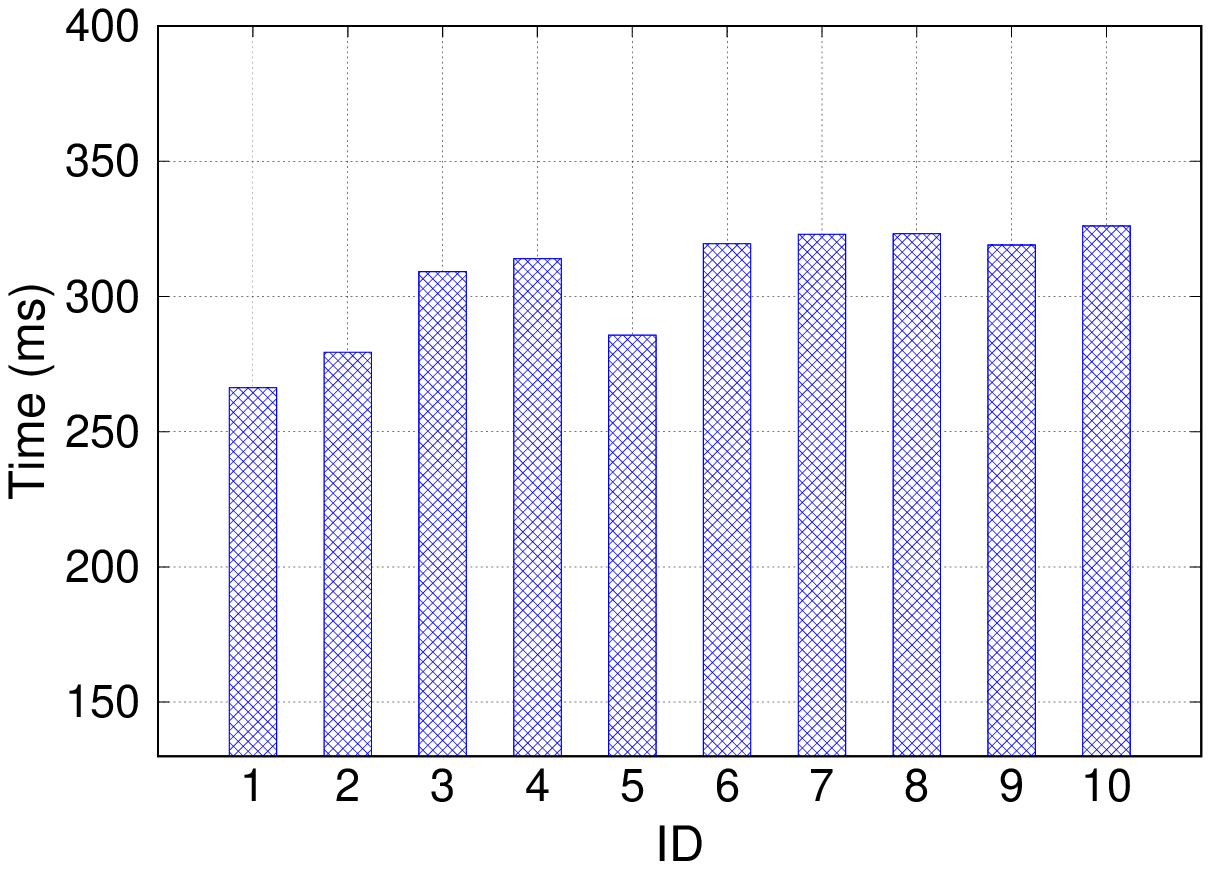}
    	\label{fig:realvictimalert}
		\subfloat[Flow installation time]
		\centering
		\includegraphics[width=0.32\linewidth]{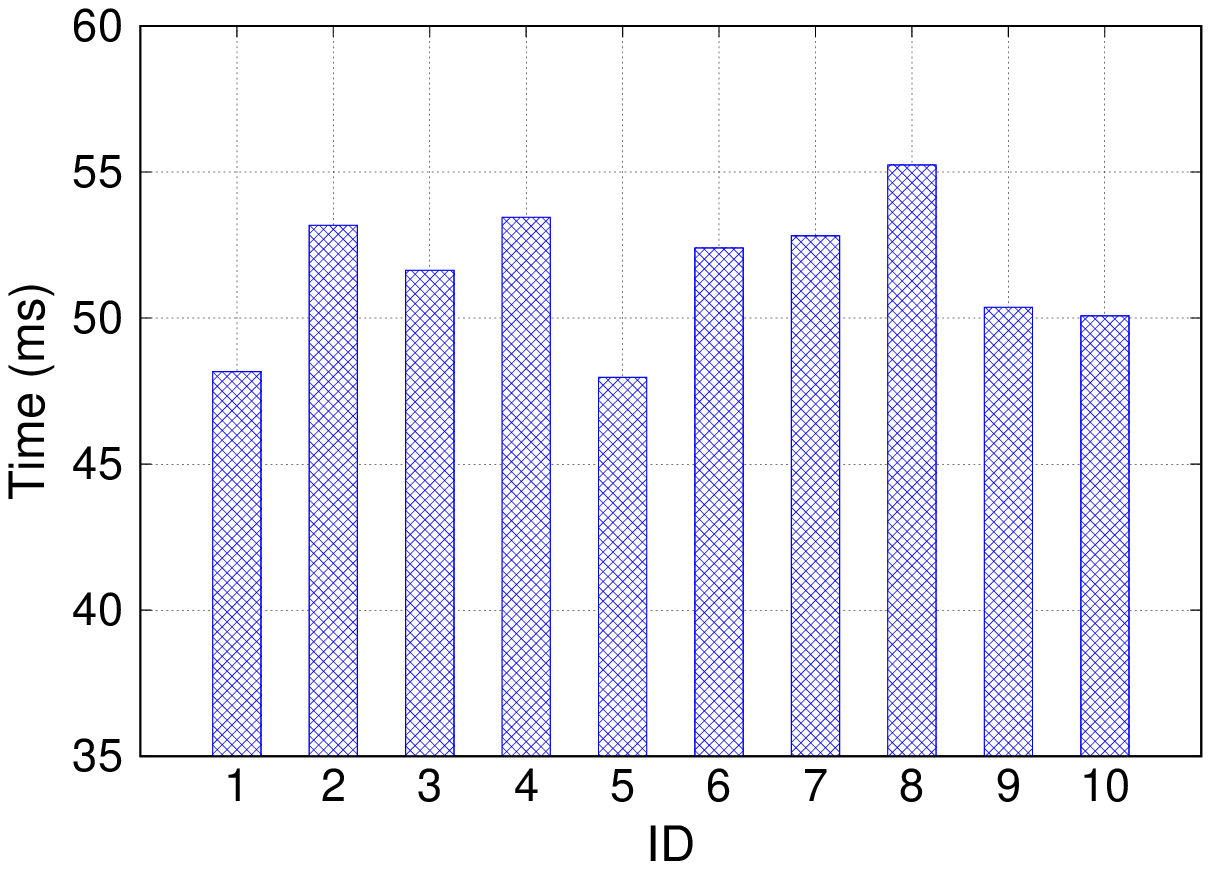}
		\label{fig:realOVSinstall}	
		\subfloat[Total time]
		\centering
		\includegraphics[width=0.32\linewidth]{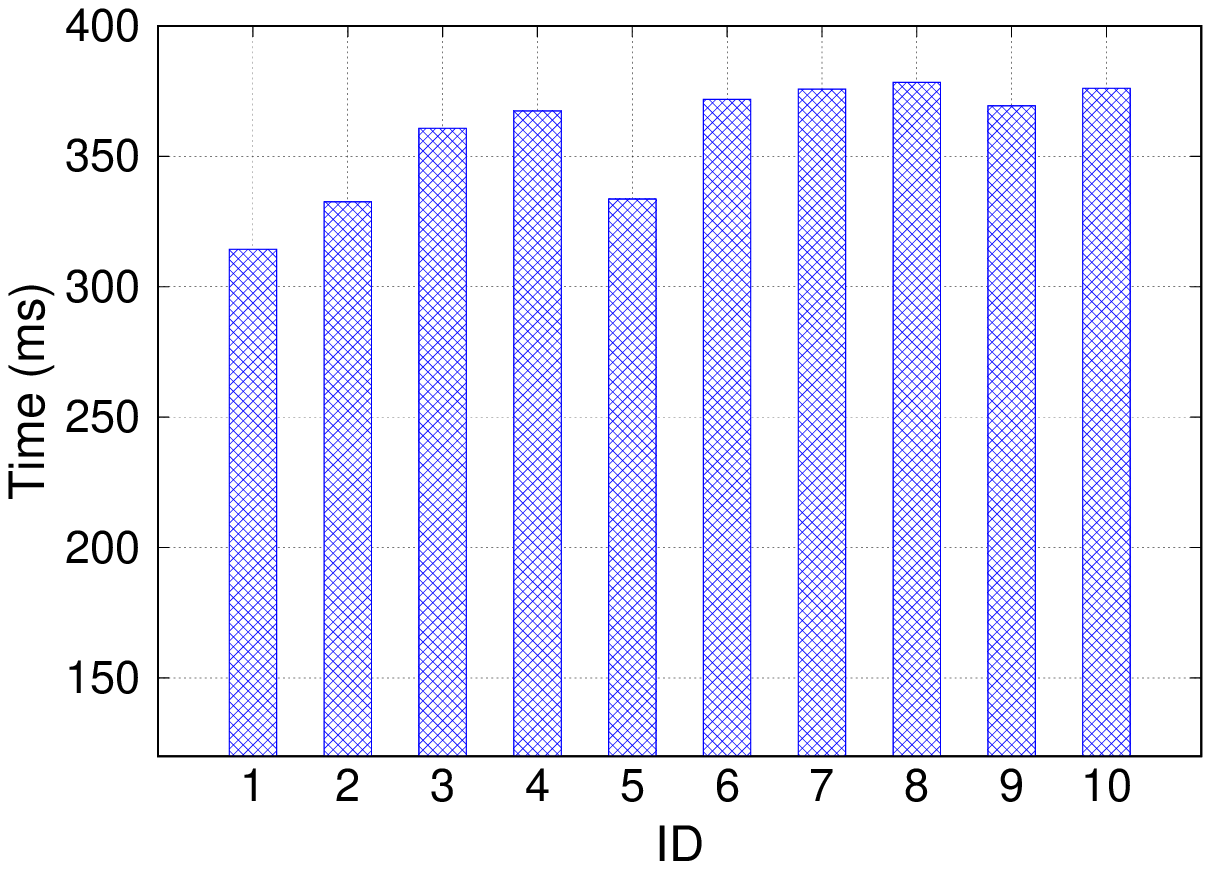}
		\label{fig:realtotalround}
	\vspace{-2mm}	
   \captionsetup{justification=centering, width=1\linewidth}
	\caption{Hardware experiments}
	\label{fig:realexperiment}
	\vspace{-8mm}
\end{figure*}


\subsubsection{Flow installation time on OpenFlow switch 1 and 2} It is challenging to obtain the flow installation time because it requires high precision time synchronization between the switch and the controller. From the controller's perspective, it can record the time when the rule was issued out to the flow queue, where rules are waiting to be installed in the flow table, but it cannot keep track of when the flow entry was successfully installed in the switch's flow table. In our experiment, we addressed this problem by the following steps. First, we set a hard timeout value for the flow entry and issue the installation command to the OpenFlow switch. When the flow entry expires, it generates an event that is handled by the controller. The timestamps of the flow issuing and the event handling procedure call can be captured at the controller. Thus, we can calculate the total round trip time from when the flow rule was issued to when the expiration message was received. In order to calculate the flow installation time, we need to subtract the hard timeout value and the single trip time from the OpenFlow switch to the controller. The single trip time can be obtained by halving the round trip time of a packet traveling from the controller to the switch and then back to the controller. Finally, Figure~\ref{fig:virtexperiment1}(b) demonstrates the flow installation time from Controller 1 to OVS-1. On average, it takes about 46$ms$ for the 10-round experiments to install a flow rule in OVS-1. Figure~\ref{fig:virtexperiment2}(b) shows similar results in OVS-2.

%
%

\subsubsection{Attacking information sharing time} It can also be called propagation or delay time from one controller to another. Figure~\ref{fig:virtexperiment2}(a) shows the time period from when Controller 1 starts sharing the attacking information to when Controller 2 accepts the information. The average sharing time is around 436$ms$. When there are multiple controllers who want to share the same attacking information, the delay determines the overall performance of the system. The delay also depends on what type of network topology being used. For example, the efficiency of a ring topology will be higher than that of a straight-line kind of topology. 

%

\subsubsection{Total time} Figure~\ref{fig:virtexperiment1}(c) shows the total time consumed to stop the malicious traffic for a single controller scenario, including alert time and flow installation time. On average, it takes 566$ms$ to block the attack in a virtual environment on GENI. Figure~\ref{fig:virtexperiment2}(c) shows the total time in a two-controller scenario, where it takes about 1003$ms$ to finish the whole loop. From the results, we can observe that the system is scalable, particularly when multiple controllers can be well connected to each other.  

%
%
%

\subsection{Results in Hardware Environment}
\subsubsection{Attack alert time}
Figure~\ref{fig:realexperiment}(a) shows the duration from when the attack was detected by the victim to the time when the controller learns about the attack from the victim. The average alert time is 306$ms$, which is less than in the virtual environment by 40\%. This can be explained by the hardware-implemented flow matching mechanism in Pica8, that operates significantly faster than the algorithm implemented in OVS.

\subsubsection{Flow installation time on Pica8 OpenFlow switch}
Figure~\ref{fig:realexperiment}(b) shows the duration between issuing the flow installation command by the controller and the actual installation on the Pica8 switch. We used similar calculation techniques as we did  in the GENI environment. It is surprising that the average time of flow installation on the hardware switch is 50$ms$, which is slower by almost 9\% than that in the OVS on GENI. The slower performance in the real environment can be explained by two main reasons: (1) Different from flow matching operations, writing on the hardware such as TCAM chips is relatively slow, because in order to guarantee fast matching, the entries on TCAM need to be sorted properly (\cite{bifulco2015towards, wen2016ruletris}); (2) The resources obtained in the GENI experiment were located on the same hardware, but in the real environment the controller and the OpenFlow switch are connected via a physical link.

\subsubsection{Total time}
Figure~\ref{fig:realexperiment}(c) shows the total time needed to detect the attack and install the rules in the Pica8 switch, so the switch will drop all subsequent packets from the attacker. The average total time is 358$ms$. The same measurement in the GENI environment is about 566$ms$ as shown in Figure~\ref{fig:virtexperiment1}(c). Overall, the performance of our solution in the hardware environment outperforms the one in the virtual environment by 37\%.  

%
%
%
%
%
%

\section{Related Work}
\label{sec:related}
The pros and cons of leveraging SDN for a network security are described by Dabbagh $et$ $al.$ in~\cite{dabbagh2015software}. In particular, the authors highlight the significance of the centralized controller that is able to detect the intrusions on a Network-Wide scale, contrary to network edge devices such as firewalls and Intrusion Detection Systems (IDS) devices. 

Skowyra $et$ $al$. in~\cite{skowyra2013software} present L-IDS, the Learning Intrusion Detection System that secures mobile devices with constrained hardware and CPU resources. In L-IDS, the SDN controller detects anomalies by collecting traffic counters values from the supervised OpenFlow switches that connect the mobile devices into one network. Moreover, the controller is able to detect physical anomalies such as unexpected relocations of the devices, a sign of a possible forge of mobile host's identity. Faluzac $et$ $al$. in~\cite{olivier2015new} present a model of an SDN-based architecture with clustered controllers for securing Internet of Things (IoT) devices. The authors show the scalability of such architecture when using border controllers that work over a specific SDN domain. In the proposed architecture, border controllers are exchanging security rules with their peers from other SDN domains, similarly to our implementation. However, paper does not elaborate on specific algorithms and shows no evaluation of an attacking scenario. The similar idea is introduced by Vandana in~\cite{vandana2016security}. In the proposed architecture, the network of IoT devices  is divided into segments with at least one SDN-capable node. Each IoT device runs an agent and registers at the closest SDN-capable node (a gateway controller). The communication between segments is provided through the gateway controllers, that authenticate the inner traffic of the IoT network and block the illegitimate traffic. As with~\cite{skowyra2013software}, the methods described in (\cite{olivier2015new, vandana2016security}) can be combined with our solution as well. In our work, we take advantage of the application information from end systems and enable the cooperation between hosts, programmable switches, and the SDN controllers, so that all IoT devices in the networks can be protected even when only one individual host was attacked by an attacker.

\section{Conclusion}
\label{sec:conclusion}


We introduce a trustworthy, cooperative and scalable architecture to enable IoT security among multiple networks. The architecture is powered by SDN technologies, where the controller application can take input about malicious activities from its end systems and translate their requirements to the network-level flow rules to stop attacks quickly. The architecture can not only benefit the victims under attacks, but any other potential targets in the networks. In addition, we solved the trust problem through double checking the traces of the malicious traffic. Meanwhile, we measured the time spent in each phase in both virtual and real environments. The results show that the overall hardware implementation outperforms the same implementation in a virtual environment. 

\bibliographystyle{IEEEtran}
\bibliography{net}
\end{document}